\PassOptionsToPackage{table}{xcolor}
\documentclass[10pt,twocolumn,letterpaper]{article}

\usepackage[pagenumbers]{cvpr} 

\usepackage{xr}

\definecolor{cvprblue}{rgb}{0.21,0.49,0.74}
\usepackage[pagebackref,breaklinks,colorlinks,allcolors=cvprblue]{hyperref}
\usepackage{amsthm}
\newtheorem{theorem}{Theorem}
\usepackage{booktabs} 
\usepackage{multirow}
\usepackage{bbm} 

\def\confName{CVPR}
\def\confYear{2027}

\title{The Fingerprinting Paradox: Why Sensitivity Undermines Black-Box Model \\ Integrity Verification}

\author{
  \textbf{Shahinul Hoque},
  \hspace{0.5cm}
  \textbf{Farhin Farhad Riya},
  \hspace{0.5cm}
  \textbf{Jinyuan Sun},
  \hspace{0.5cm}
  \textbf{Fnu Suya} \\
  University of Tennessee, Knoxville\\
  \texttt{shoque@vols.utk.edu,
  friya@vols.utk.edu,
  jysun@utk.edu,
  fsuya@utk.edu}
}

\begin{document}
\maketitle


\begin{abstract}
Black-box integrity verification aims to detect unauthorized modifications to outsourced models using only prediction APIs. Existing frameworks seek to improve verification efficiency through \emph{sensitive fingerprints}: queries whose predictions are more likely to change under tampering than those of genuine inputs. We show that this sensitivity can also help dishonest providers recognize verification queries and selectively conceal tampering. We prove that excess sensitivity relative to genuine inputs lower-bounds statistical distinguishability. We evaluate five integrity verification frameworks across three model architectures, three datasets, and three adversarial knowledge settings. When the deployed framework is known, our fingerprint detectors achieve AUCs at least 0.989, and selective filtering makes all five frameworks require significantly larger estimated query budgets than genuine inputs, reversing their intended efficiency advantage. A query-efficiency advantage remains marginally in a small subset of evaluated settings with poor detection transfer. We further design verification methods under stronger concealment constraints but find that reduced detectability comes with losses in sensitivity or fingerprint-generation success. Our results expose a sensitivity--detectability tension and show that sensitivity alone does not ensure efficient black-box verification against adaptive malicious providers.

\end{abstract}


\section{Introduction}


Model owners outsource inference to third-party providers to reduce deployment costs and scale their services~\cite{aws_sagemaker_inference}. However, owners must verify that their models are deployed without unauthorized modifications. Providers control model execution, while owners may have only API access and cannot directly inspect the models. Black-box integrity verification therefore seeks to detect modifications through API responses alone. We study the label-only setting, in which the API returns only prediction labels.

Existing integrity verification frameworks (IVFs) address this problem by constructing sensitive fingerprints: inputs whose predictions are more likely to change under model modifications than those of genuine inputs~\cite{xiaofan2024intersecting,bai2025sdbf,he2019sensitive,wang2023publiccheck,he2024towards}. The owner records the original model's predictions on these fingerprints and compares them with the provider's responses. By concentrating verification on sensitive inputs, IVFs aim to expose tampering with fewer queries than genuine inputs.

This efficiency, however, depends on fingerprints remaining \emph{difficult} for the provider to recognize. A provider that identifies verification queries can selectively return predictions from a retained copy of the original model while serving a modified model to ordinary users. Keeping the exact fingerprints secret does not necessarily hide the statistical characteristics shared by independently generated fingerprints. In particular, optimizing fingerprints to respond differently from genuine inputs under model modifications may itself provide a signal for recognizing them. We therefore ask whether sensitive fingerprints retain their verification advantage against a provider that actively tries to identify and selectively answer verification queries.

\begin{figure*} [htbp]
    \centering
    \includegraphics[width=0.95\linewidth]{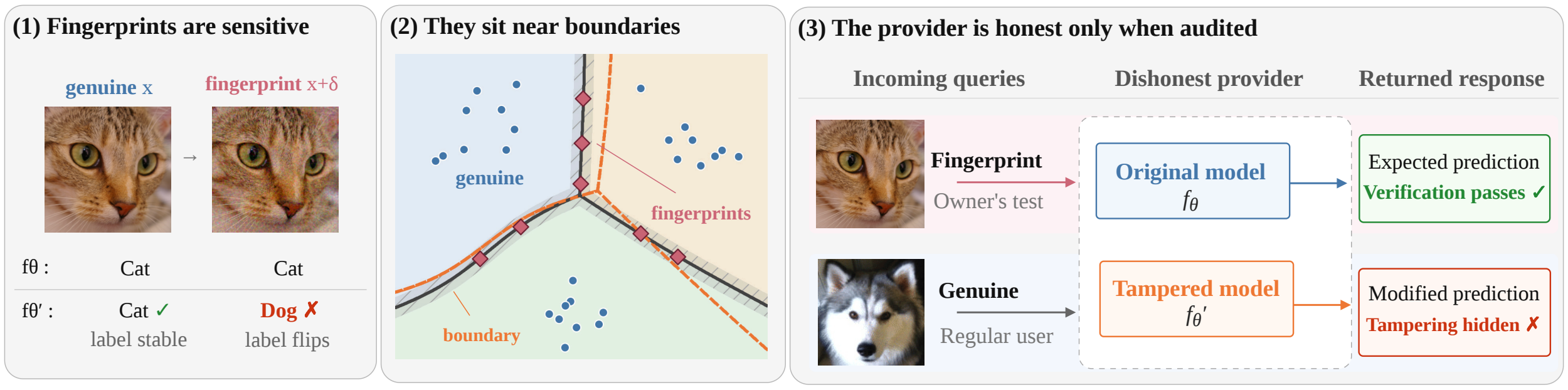}
    \caption{Overview of the fingerprint detectability threat. Here, $f_\theta$ denotes the original and $f_{\theta}'$ denotes the deployed tampered model.}
    \label{fig:overview}
\end{figure*}

\noindent\textbf{Contributions.}
We first theoretically show that excess fingerprint sensitivity to model modifications for verification purposes lower-bounds their statistical distinguishability from genuine inputs (Section~\ref{sec:theory_bound}). To determine whether a dishonest provider can exploit this distinction, we develop detectors using features from a single model forward pass under three adversarial knowledge settings (Section~\ref{sec:system_design}) and evaluate them across five IVFs, three architectures, and three datasets, obtaining average AUCs of 0.989--1.000 when the deployed IVF is known (Section \ref{sec:evaluation}). 

Our selective filtering makes all five evaluated IVFs require substantially more queries than genuine inputs to obtain the same number of label disagreements at the same target success probability when the IVF is known, reversing the query-efficiency advantage these IVFs were designed to provide. A query-efficiency advantage remains only in a small subset of the evaluated transfer settings, where detectors trained on one IVF generalize poorly to a substantially different IVF (Section~\ref{sec:defense1}). To improve concealment even against a provider that knows the framework, we examine two approaches: tightening MiSentry's existing entropy constraint and extending its generation procedure with output-margin and representation constraints through MAZE. Both reduce detectability but incur losses in sensitivity or generation success, with MAZE providing little average sensitivity advantage and query costs close to those of genuine inputs (Section~\ref{sec:defense2}).

These findings provide empirical evidence of the sensitivity--detectability tension and motivate future IVFs that preserve verification efficiency against providers that actively identify and filter verification queries.




\section{Background and Threat Model}
We first introduce background on fingerprint-based black-box model integrity verification (Section~\ref{sub:verification}) and provide the threat model (Section~\ref{sub:threat_model}).

\subsection{Black-Box Integrity Verification}
\label{sub:verification}

In model outsourcing, a model owner deploys a trained model on a third-party service provider's infrastructure. A dishonest provider may modify the model to introduce malicious behavior. We consider the black-box verification setting in which the owner can query the deployed model through an API but receives only its top-1 predictions~\cite{xiaofan2024intersecting,bai2025sdbf,he2024towards,wang2023publiccheck}.


\paragraph{Notations.} Let $f_\theta$ and $f_{\theta}'$ denote the original and deployed models, respectively. For an input $x$, let $\mathbf{y}$ and $\mathbf{p}=\operatorname{softmax}(\mathbf{y})$ denote the original model's logits and class probabilities, with $\hat{y}_\theta(x)=\arg\max_c p_c$ denoting its predicted label. Integrity verification frameworks (IVFs) construct \emph{sensitive fingerprints} whose predictions are likely to change under model modifications. The verifier records the original model's predictions and compares them with $\hat{y}_{\theta'}(x)$ returned by the deployed model. Prediction disagreements provide evidence of potential tampering.

Fingerprint generation keeps the original model fixed and searches for \emph{sensitive} inputs that expose changes to its behavior. IBSF, SDBF, SSF, and MiSentry optimize pixel-space perturbations of genuine images, whereas PublicCheck searches the latent space of a generative model. Their objectives seek sensitivity in different ways.

\noindent\textbf{IBSF}~\cite{xiaofan2024intersecting} generates fingerprints near $K$-class decision-boundary intersections by maximizing partial Shannon entropy $-\sum_{c\in\mathcal{T}_K}p_c\log p_c$, where $\mathcal{T}_K$ contains the top-$K$ classes without probability renormalization. For fixed $K\geq 3$, this encourages probabilities near $1/K$, approaching boundary intersections where sensitivity increases with $K$.

\noindent\textbf{SDBF}~\cite{bai2025sdbf} generates fingerprints near steep decision boundaries by minimizing $\mathrm{var}(\mathbf{y}_{C_{\mathrm{sub}}})-\lambda\|\mathbf{y}_{C_{\mathrm{sub}}}\|_2$, where $C_{\mathrm{sub}}$ contains the $K$ top-scoring classes and $\lambda$ balances the terms. Minimizing variance encourages logit equality, while maximizing the logit magnitude aims to increase sensitivity.

\noindent\textbf{PublicCheck}~\cite{wang2023publiccheck} generates fingerprints near decision boundaries by perturbing VQ-VAE latent codes of genuine images until the decoded images' top-1 predictions change. Samples near these transitions serve as fingerprints sensitive to boundary shifts caused by model modifications.

\noindent\textbf{SSF}~\cite{he2019sensitive} generates fingerprints by maximizing the sensitivity of model outputs to small perturbations in the final-layer weights. It optimizes the gradient magnitude of an aggregated log-probability statistic over a fixed subset of classes, without explicitly locating decision boundaries.

\noindent\textbf{MiSentry}~\cite{he2024towards} generates fingerprints using meta-learning to maximize prediction divergence between the original model and a zoo of modified variants, without explicitly targeting decision boundaries. To improve covertness, it constrains output entropy to a range derived from genuine inputs.

Figure~\ref{fig:overview} shows an overview of the threat, where a provider that recognizes fingerprint queries can selectively return the original model's predictions for those queries while serving a modified model to ordinary users. Effective verification therefore also requires fingerprints that are difficult to distinguish from genuine queries. We investigate whether the properties used to make fingerprints sensitive also allow a dishonest provider to identify them.

\subsection{Threat Model}
\label{sub:threat_model}
\paragraph{Setting.} We consider a model-outsourcing setting in which a model owner supplies a trained model (with full access to it) to a third-party service provider and wants to verify that the provider deploys it without unauthorized modifications. The owner or a trusted auditor performs verification by querying the service API and observing only the returned top-1 predictions. The provider hosts the model and acts as the adversary. We assume that the verifier and its private fingerprint storage are trusted.

\paragraph{Adversarial objective, knowledge, and capabilities.}
The adversary aims to serve a modified model to genuine users while evading integrity verification. It has white-box access to the hosted model. Prior IVFs allow the adversary to know the fingerprint-generation algorithm. Therefore, we study both this setting and settings in which the deployed algorithm is unknown, as detailed in Section~\ref{sec:system_design}. The adversary can inspect incoming queries, control how they are processed, and choose the predictions returned through the API. It may also retain a copy of the original model before modification. The verifier's selected fingerprints are not disclosed to the adversary before verification. We exclude adversaries who collude with the trusted verifier or compromise the stored secret fingerprints.





\section{Sensitivity-Detectability Analysis}
\label{sec:theory_bound}
Current IVFs make fingerprints more sensitive to model modifications than genuine inputs, enabling verification with few queries. We hypothesize that this difference in sensitivity also makes fingerprint queries statistically distinguishable from genuine queries.

To study distinguishability across queries, let $P_F$ and $P_G$ denote the fingerprint and genuine-input distributions, respectively. Sensitivity of fingerprinting samples also depends on the model modification being tested: an input may change its predicted label under one modification but retain it under another. We therefore measure sensitivity over a specified distribution $T$ of modified model parameters $\theta'$. We define an input's tampering sensitivity $s_T(x)$ as its probability of changing labels under a modification drawn from $T$, and the excess sensitivity $\Delta_T$ as the difference in expected sensitivity between fingerprints and genuine inputs:
\begin{equation}
\begin{aligned}
s_T(x) &= \Pr_{\theta'\sim T}
[\hat{y}_{\theta'}(x)\neq\hat{y}_\theta(x)], \\
\Delta_T &= \mathbb{E}_{x\sim P_F}[s_T(x)]
- \mathbb{E}_{x\sim P_G}[s_T(x)],
\end{aligned}
\label{eq:sensitivity}
\end{equation}
where $\hat{y}_\theta(x)$ denotes the original model's top-1 prediction. A positive $\Delta_T$ indicates that fingerprints change labels more often than genuine inputs under the same modification distribution. The following theorem quantifies the statistical distinguishability implied by this sensitivity gap.

\begin{theorem}[Sensitivity--Detectability Lower Bound]
\label{thm:bound}
For any tamper distribution $T$,
\begin{equation}
\begin{aligned}
\mathrm{TV}(P_F,P_G) &\geq |\Delta_T|, \quad
\mathrm{BA}_{\mathrm{opt}} &\geq
\frac{1+|\Delta_T|}{2},
\end{aligned}
\label{eq:bound}
\end{equation}
where $\mathrm{TV}$ denotes total variation distance and
$\mathrm{BA}_{\mathrm{opt}}$ is the optimal balanced accuracy (BA) for distinguishing fingerprints from genuine inputs. BA averages the true-positive and true-negative rates, giving equal weight to fingerprints and genuine inputs regardless of their relative prevalence.\end{theorem}

The proof of Theorem~\ref{thm:bound}, the construction of a sensitivity detector, and details on the model modification distribution $T$ are provided in Appendix~\ref{app:proof}.

\paragraph{Implication.}
The theorem shows that the sensitivity advantage sought by IVFs can itself reveal verification queries. When $\Delta_T>0$, a provider can sample a modified model from $T$ and classify a query as a fingerprint if its predicted label differs from that of the original model. This rule achieves an expected BA of $(1+\Delta_T)/2$ without access to the verifier's secret fingerprints. 

Below, we empirically estimate $\Delta_T$ using 30 modified models per model--dataset configuration: five models from each of six modification families (last-layer fine-tuning \cite{kumar2022fine}, targeted attack \cite{shafahi2018poison}, BadNets backdoor \cite{gu2019badnets}, bit-flip \cite{rakin2019bit}, weight quantization \cite{jacob2018quantization}, and magnitude pruning \cite{han2015learning}). This produces 270 modified models across the nine configurations. For each configuration, we measure label-flip rates under these models for 500 fingerprints per IVF method described in Section \ref{sub:verification} and 500 genuine samples.

\begin{table}[ht]
\centering
\scriptsize
\setlength{\tabcolsep}{7pt}
\caption{Estimated sensitivity--detectability bounds in our experimental setting, averaged over nine configurations. BA (\%) is the theoretical lower bound on optimal balanced accuracy.}
\label{tab:sensitivity_bound}
\begin{tabular}{lccccc}
\toprule
Metric & IBSF & SDBF & SSF & PublicCheck & MiSentry \\
\midrule
$\widehat{\Delta}_T$ & 0.387 & 0.423 & 0.059 & 0.223 & 0.137 \\
BA (\%) & 69.3 & 71.1 & 52.9 & 61.1 & 56.9 \\
\bottomrule
\end{tabular}
\end{table}

Table~\ref{tab:sensitivity_bound} shows that the estimated sensitivity gap is positive for all five IVFs. Averaged over the nine configurations, the corresponding estimated lower bound on optimal BA ranges from 52.9\% (SSF) to 71.1\% (SDBF). The bound depends only on the sensitivity gap, so these values estimate the distinguishability that this statistic alone guarantees under the evaluated model modifications. The following sections develop practical detectors that exploit additional fingerprint characteristics to achieve substantially higher detection accuracy.


\section{Fingerprint Detection}
\label{sec:system_design}

Detecting fingerprints through prediction disagreement is costly because the provider must evaluate multiple model variants for each query. We instead train a binary classifier on logits and intermediate representations from a single forward pass, combined with statistics computed directly from the input. At detection time, this approach requires only feature computation and classifier inference. It needs no additional model variants and no gradient computation.

We consider three settings. In \emph{framework-specific detection} (Section~\ref{sub:eval_framework_specific}), the provider knows the deployed IVF. In \emph{ensemble detection} (Section~\ref{sub:eval_ensemble}), the provider does not know which IVF is deployed, but detector training includes fingerprints from that framework. In \emph{cross-framework generalization} (Section~\ref{sub:eval_generalizability}), fingerprints from the evaluated IVF are excluded from detector training.

\subsection{Framework-Specific Detectors}
\label{sub:framework_spec}

We first consider an adversary who knows the deployed IVF and can therefore train a framework-specific detector, consistent with the adversarial knowledge assumed by prior IVFs. Drawing on recent out-of-distribution (OOD) detection literature~\cite{hendrycks2016baseline,lee2018simple,wang2022vim,liu2020energy,sun2021react}, we combine general anomaly-detection features, including output confidence, logit statistics, Mahalanobis distances, nearest-neighbor distances, and PCA reconstruction residuals, with a small set of IVF-specific features. IBSF drives the top-$K$ class probabilities toward a multiclass boundary intersection, so we characterize the resulting top-$K$ probability and logit profile. SDBF explicitly reduces variance among selected logits while increasing their magnitude, motivating features that measure top-$K$ logit equality, magnitude, and separation. SSF maximizes sensitivity to final-layer weight perturbations, so we derive closed-form sensitivity measures from the logits and penultimate activations. PublicCheck perturbs VQ-VAE latent codes until the predicted label changes, motivating decision-boundary margins together with image statistics that can capture effects of latent-space reconstruction. MiSentry constrains output entropy to a genuine-input-derived interval while maximizing disagreement with modified models, so we measure entropy-band conformity together with general output and representation anomalies. Only the PublicCheck detector uses image statistics, a \emph{general} description that is not tied to any specific IVF, to capture potential artifacts introduced by its generative, VQ-VAE-based fingerprint construction. Appendix \ref{app:features} provides the complete feature table for each framework-specific detector.

Using the extracted features, we train a 500-tree random forest for each IVF, with a minimum leaf size of two. Random forests accommodate heterogeneous output, representation, and objective-specific features and capture potentially nonlinear relationships with fingerprint detectability. To reflect the expected class imbalance in deployment, we train with more genuine samples than fingerprints. We select each detector's threshold by maximizing Youden's $J$ statistic~\cite{fluss2005estimation}, $\mathrm{TPR}-\mathrm{FPR}$, on out-of-bag predictions. Since $J$ uses class-normalized rates, threshold selection is not dominated by the larger genuine sample set.

\begin{table*}[tbp]
\centering
\scriptsize
\setlength{\tabcolsep}{3pt}
\renewcommand{\arraystretch}{0.95}
\caption{Fingerprint detection per model architecture under three adversary settings. Recall, FPR, and BA are in \%; each architecture block is averaged over CIFAR-100, GTSRB, and ImageNet-1K, and the last block over all nine configurations. Ensemble detectors use a single calibrated threshold, so their FPR is shared across all five IVF ensemble. PC: PublicCheck; MS: MiSentry.}
\label{tab:eval_main}
\begin{tabular*}{\textwidth}{@{\extracolsep{\fill}}ll |ccccc|ccccc|ccccc}
\toprule
\multirow{2}{*}{Model} & \multirow{2}{*}{Metric}
 & \multicolumn{5}{c|}{(a) Framework-specific}
 & \multicolumn{5}{c|}{(b) Ensemble (all five IVFs)}
 & \multicolumn{5}{c}{(c) LOFO (IVF held out)} \\
\cmidrule(lr){3-7}\cmidrule(lr){8-12}\cmidrule(lr){13-17}
 & & IBSF & SDBF & SSF & PC & MS
   & IBSF & SDBF & SSF & PC & MS
   & IBSF & SDBF & SSF & PC & MS \\
\midrule
\rowcolor{gray!15}
 \cellcolor{white} & Recall &  99.9 & 100.0 &  99.7 &  97.1 &  94.3 & 100.0 & 100.0 & 100.0 &  98.9 &  92.9 & 100.0 & 100.0 & 100.0 &  62.7 &  78.4 \\
 & FPR  &   0.0 &   0.0 &   0.1 &   0.6 &   4.9 &   3.2 &   3.2 &   3.2 &   3.2 &   3.2 &   3.8 &   3.7 &   3.8 &   3.1 &   3.1 \\
 & BA   &  99.9 & 100.0 &  99.9 &  98.2 &  94.7 &  98.4 &  98.4 &  98.4 &  97.9 &  94.9 &  98.1 &  98.2 &  98.1 &  79.8 &  87.7 \\
 \multirow{-4}{*}{ResNet-50} & AUC  & 1.000 & 1.000 & 1.000 & 0.999 & 0.988 & 1.000 & 1.000 & 1.000 & 0.998 & 0.991 & 1.000 & 1.000 & 1.000 & 0.905 & 0.947 \\
\addlinespace[2pt]
\rowcolor{gray!15}
 \cellcolor{white} & Recall &  99.9 & 100.0 &  99.9 &  98.8 &  98.5 & 100.0 & 100.0 & 100.0 &  98.2 &  99.0 & 100.0 & 100.0 & 100.0 &  50.7 &  88.4 \\
 & FPR  &   0.0 &   0.0 &   0.0 &   1.5 &   3.3 &   3.1 &   3.1 &   3.1 &   3.1 &   3.1 &   3.1 &   3.3 &   3.1 &   2.7 &   2.8 \\
 & BA   & 100.0 & 100.0 &  99.9 &  98.7 &  97.6 &  98.5 &  98.5 &  98.5 &  97.6 &  98.0 &  98.5 &  98.4 &  98.5 &  74.0 &  92.8 \\
 \multirow{-4}{*}{MobileNetV3} & AUC  & 1.000 & 1.000 & 1.000 & 0.999 & 0.996 & 1.000 & 1.000 & 1.000 & 0.997 & 0.997 & 1.000 & 1.000 & 1.000 & 0.850 & 0.985 \\
\addlinespace[2pt]
\rowcolor{gray!15}
 \cellcolor{white} & Recall &  99.9 & 100.0 &  99.7 &  98.4 &  95.5 & 100.0 & 100.0 & 100.0 &  99.2 &  80.4 & 100.0 & 100.0 &  94.1 &  49.8 &  30.4 \\
 & FPR  &   0.0 &   0.0 &   0.2 &   0.9 &   6.3 &   3.9 &   3.9 &   3.9 &   3.9 &   3.9 &   4.4 &   4.4 &   3.9 &   3.6 &   3.6 \\
 & BA   &  99.9 & 100.0 &  99.8 &  98.8 &  94.6 &  98.1 &  98.1 &  98.1 &  97.6 &  88.3 &  97.8 &  97.8 &  95.1 &  73.1 &  63.4 \\
 \multirow{-4}{*}{ViT-B/16} & AUC  & 1.000 & 1.000 & 1.000 & 1.000 & 0.983 & 0.999 & 0.999 & 0.999 & 0.996 & 0.970 & 0.999 & 0.999 & 0.976 & 0.913 & 0.874 \\
\midrule
\rowcolor{gray!15}
 \cellcolor{white} & \textbf{Recall} & \textbf{99.9} & \textbf{100.0} & \textbf{99.8} & \textbf{98.1} & \textbf{96.1} & \textbf{100.0} & \textbf{100.0} & \textbf{100.0} & \textbf{98.7} & \textbf{90.8} & \textbf{100.0} & \textbf{100.0} & \textbf{98.0} & \textbf{54.4} & \textbf{65.7} \\
 & FPR  &   0.0 &   0.0 &   0.1 &   1.0 &   4.8 &   3.3 &   3.3 &   3.3 &   3.3 &   3.3 &   3.8 &   3.8 &   3.6 &   3.1 &   3.1 \\
 & BA   &  99.9 & 100.0 &  99.8 &  98.6 &  95.6 &  98.3 &  98.3 &  98.3 &  97.7 &  93.7 &  98.1 &  98.1 &  97.2 &  75.6 &  81.3 \\
 \multirow{-4}{*}{\textbf{Average}} & AUC  & 1.000 & 1.000 & 1.000 & 0.999 & 0.989 & 0.999 & 0.999 & 0.999 & 0.997 & 0.986 & 0.999 & 0.999 & 0.992 & 0.889 & 0.935 \\
\bottomrule
\end{tabular*}
\end{table*}

\subsection{Ensemble Detector}
\label{sub:ensemble}
We use an ensemble detector when the provider does not know which IVF is deployed. The same ensemble construction supports two settings, ensemble detection in Section \ref{sub:eval_ensemble} and the cross-framework generalization in Section \ref{sub:eval_generalizability}.

To avoid relying on knowledge of the deployed IVF's generation objective or settings, we exclude objective-specific features, such as MiSentry's entropy-band margin and SSF's sensitivity objective. We retain the remaining features from Section \ref{sub:framework_spec}, including the image statistics for all IVFs. These statistics describe generic input properties and may reveal fingerprint artifacts without requiring knowledge of how the fingerprints were generated. All evaluations use this feature set with details in Appendix~\ref{app:features}.

We initially trained a single random forest on fingerprints from all IVFs but observed poor detection performance due to inadequate threshold calibration. We therefore train a separate expert for each \emph{available} IVF to capture framework-specific fingerprint characteristics and enable independent threshold calibration. Each expert is a 500-tree random forest with a minimum leaf size of two, trained using class-balanced weights on fingerprints from its corresponding IVF and a shared, larger set of genuine inputs. We calibrate each expert's score using its empirical percentile $q_i(x)$ in the genuine out-of-bag score distribution, which performed best among the aggregation strategies evaluated. A query is flagged when $\max_i q_i(x)>1-\alpha/K$, where $K$ is the number of experts and $\alpha$ specifies the target false-positive rate (we set $\alpha$=0.05 throughout our experiments). The Bonferroni correction controls the aggregate false-positive rate across experts. We use the same ensemble detection procedure across all evaluation settings.


\section{Evaluating Fingerprint Detectability}
\label{sec:evaluation}

\paragraph{Experimental setup.} Following the evaluation settings in prior IVFs, we evaluate the detectors from Section~\ref{sec:system_design} across nine model--dataset configurations, combining ResNet-50~\cite{resnet50}, MobileNetV3~\cite{mobilenetv3}, and ViT-B/16~\cite{vit_b16} models with CIFAR-100~\cite{cifar100}, GTSRB~\cite{gtsrb}, and ImageNet-1K~\cite{imagenet} dataset. We report recall, false positive rate (FPR), balanced accuracy (BA), as defined in Theorem \ref{thm:bound}, and area under the curve (AUC). We emphasize recall because identifying verification queries allows the adversary to conceal model tampering by selectively returning the original model's predictions. We evaluate on fingerprints from generation runs held out from both training and threshold selection. Further experimental details are provided in Appendix~\ref{app:setup}. 

\subsection{Framework-Specific Fingerprint Detection}
\label{sub:eval_framework_specific}

Table~\ref{tab:eval_main}(a) reports framework-specific detection on held-out fingerprint runs, averaging results across the three datasets for each model architecture. All five IVFs are highly detectable. SDBF achieves 100\% recall with zero FPR on every architecture, while IBSF and SSF achieve average recalls of 99.9\% and 99.8\%, respectively, with FPR of at most 0.2\%. PublicCheck also remains highly detectable, achieving 98.1\% recall at 1.0\% FPR despite generating fingerprints in latent space. MiSentry is comparatively harder to detect, with 96.1\% recall at 4.8\% FPR. Its detectability is also the least uniform: recall ranges from 94.3\% on ResNet-50 to 98.5\% on MobileNetV3, FPR rises to 6.3\% on ViT-B/16, and recall drops to 91.6\% on ResNet-50-CIFAR-100 in the per-configuration results (Appendix~\ref{app:eval_specific}). Overall, MiSentry's explicit entropy constraint does not remove the detectable differences in output and representation statistics.

\subsection{Ensemble Fingerprint Detection}
\label{sub:eval_ensemble}

Table~\ref{tab:eval_main}(b) reports ensemble detection when the deployed IVF is unknown but included in detector training. Without framework knowledge, the ensemble reaches 100\% average recall on IBSF, SDBF, and SSF, 98.7\% on PublicCheck, and 90.8\% on MiSentry. The trained ensemble detector gives an average FPR of 3.3\%, BA between 93.7\% and 98.3\%, and AUC of at least 0.986 for every IVF. Recall on SSF and PublicCheck is slightly higher than with the framework-specific detectors (99.8\% and 98.1\% in Table~\ref{tab:eval_main}(a)), which suggests that different IVFs share detectable characteristics the ensemble can exploit. MiSentry remains the hardest case, and its difficulty is architecture-dependent: recall is 99.0\% on MobileNetV3 and 92.9\% on ResNet-50 but 80.4\% on ViT-B/16 (AUC 0.970), with per-configuration recall as low as 67.8\% on ViT-B/16/ImageNet-1K (Appendix~\ref{app:eval_ensemble}). We hypothesize that MiSentry's entropy constraint suppresses distinctive output signatures and that ViT's patch-based representations further weaken perturbation-induced features. But we do not isolate these two factors here. Overall, a detector trained on multiple IVFs can reliably identify verification queries without knowing which framework is deployed. The cost is a higher FPR than the framework-specific detectors for all IVFs except MiSentry, but it is still a cost the provider can easily pay.

\subsection{Generalizability of Fingerprint Detection}
\label{sub:eval_generalizability}

We evaluate generalization to IVFs excluded from detector training in two settings. For leave-one-framework-out (LOFO) detection, we exclude the target IVF, train framework-specific experts on the remaining four IVFs, and combine their genuine-calibrated scores as described in Section~\ref{sub:ensemble}. For single-source transfer, we train a framework-specific detector on one IVF and evaluate it directly on each of the other four IVFs to measure pairwise transfer of detection signals. These represent the permissive and restrictive settings of testing generalization.

\paragraph{Leave-one-framework-out (LOFO) detection.} Table~\ref{tab:eval_main}(c) reports the LOFO results. The gradient-based IVFs (IBSF, SDBF, and SSF) remain highly detectable despite being excluded from training, with average recall of 100\% for IBSF and SDBF and 98.0\% for SSF. On average, BA is above 97\% and AUC is above 0.99. In contrast, recall falls to 54.4\% for PublicCheck and 65.7\% for MiSentry, with BA of 75.6\% and 81.3\%, respectively. Generalization also varies by architecture: PublicCheck recall remains between 49.8 and 62.7\%, while MiSentry drops from 78.4\% and 88.4\% on ResNet-50 and MobileNetV3 to 30.4\% on ViT-B/16. Per-configuration results in Appendix~\ref{app:eval_generalization} show the weakest cases on ImageNet-1K, where PublicCheck recall falls below 17\%. These results suggest that the gradient-based IVFs share transferable detection signatures, whereas PublicCheck and MiSentry contain more framework-specific characteristics not shared by others.

\begin{figure}[hbt]
\centering
\includegraphics[width=\columnwidth]{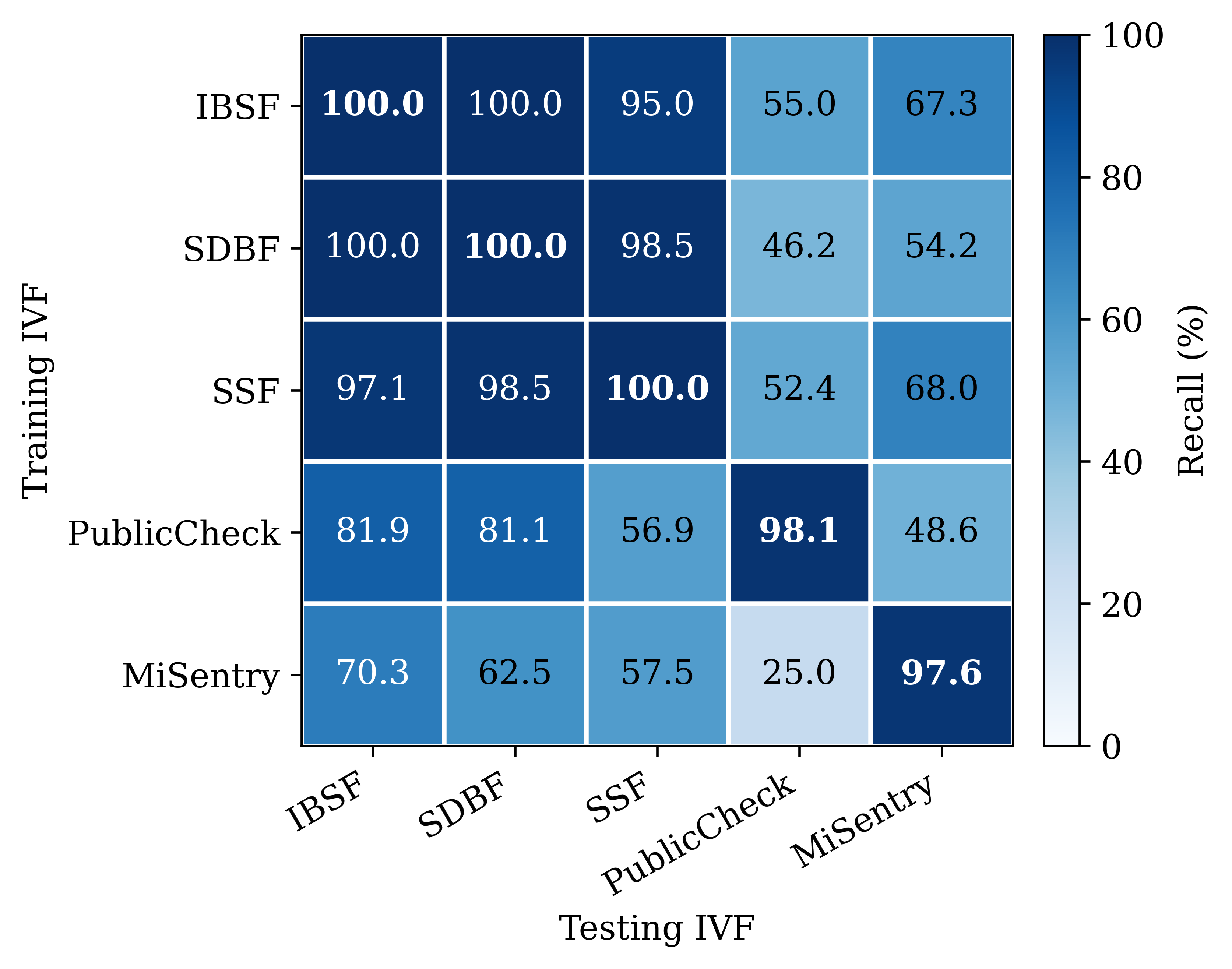}
\caption{Cross-IVF transfer of framework-specific detectors (recall, \%), averaged over nine model--dataset configurations. The Mean row and column average the off-diagonal entries.}
\label{fig:transfer_5x5}
\vspace{-0.4cm}
\end{figure}

\paragraph{Single-source transfer.} Figure~\ref{fig:transfer_5x5} reports recall averaged over the nine model-dataset pairs. Consistent with the LOFO results, detectors trained on gradient-based methods (IBSF, SDBF, or SSF) achieve 95.0--100\% recall on the other gradient-based IVFs, with perfect transfer between IBSF and SDBF. Transfer to PublicCheck and MiSentry is substantially weaker, reaching 46.2--55.0\% and 54.2--68.0\% recall, respectively. Transfer is also asymmetric: a PublicCheck-trained detector identifies 81.9\% of IBSF and 81.1\% of SDBF fingerprints, whereas the reverse directions reach only 55.0\% and 46.2\%. We hypothesize that this asymmetry reflects their generation objectives: PublicCheck targets boundaries between competing classes and may therefore capture characteristics also induced by the multiclass-boundary objectives of IBSF and SDBF. MiSentry shows the weakest transfer to PublicCheck (25.0\%), reflecting differences in their generation objectives. Overall, cross-framework detection is strongest when IVFs share similar statistical signatures.

\section{Impact on Query Efficiency after Filtering}
Having shown that a dishonest provider can identify verification queries, we next examine the consequences for verification efficiency. We first quantify how selective filtering changes the query budgets required by existing IVFs, using genuine inputs as a baseline (Section~\ref{sec:defense1}). To recover the efficiency lost to filtering, we then investigate whether modifying fingerprint generation can reduce detectability while preserving tampering sensitivity (Section~\ref{sec:defense2}). The resulting query costs and trade-offs in fingerprint generation illustrate the practical implications of the sensitivity--detectability tension characterized in Theorem~\ref{thm:bound}.

\subsection{Query Efficiency Drops for Existing IVFs}
\label{sec:defense1}

A dishonest provider can conceal tampering by returning the original model's predictions for queries identified as fingerprints. Consequently, only fingerprints that evade detection and produce a label disagreement can expose the modification. Filtering therefore increases the query budget required to obtain the same number of label disagreements at the same target success probability. We quantify this cost for a fixed modified model. Let $d$ denote the detector's recall, the probability that a fingerprint is filtered, and let $S_F$ denote the probability that a fingerprint's predicted label changes under the modification. If filtering and label changes are independent, the probability that a fingerprint both evades detection and reveals tampering is $p_{\mathrm{ideal}}=(1-d)S_F$. More generally, for $n$ independently sampled fingerprints, the number of observed disagreements follows $X\sim\mathrm{Binomial}(n,p)$, where $p$ is the joint probability of evading detection and producing a label change. To observe at least $m$ disagreements with probability at least $1-\delta$, the verifier requires the smallest $n$ satisfying $\Pr[X\ge m]\ge 1-\delta$. Under the independence assumption, this query requirement increases as $d$ increases.

\begin{table}[ht]
\centering
\footnotesize
\setlength{\tabcolsep}{4pt}
\caption{Effect of fingerprint filtering on tampering detection, pooled over nine configurations and 30 tampered models per configuration. $N=135{,}000$ denotes input--model pairs per IVF. $p_{\mathrm{bef}}$ and $p_{\mathrm{aft}}$ are measured flip probabilities before and after filtering.}
\label{tab:survivors}
\resizebox{\columnwidth}{!}{
\begin{tabular}{lcccccc}
\toprule
IVF & $p_{\mathrm{bef}}$ (\%) & $p_{\mathrm{aft}}$ (\%) & $p_{\mathrm{ideal}}$ (\%) & $d$ (\%) & $n_{\mathbb{E}}$ & $n_{95}$ \\
\midrule
IBSF        & 45.76 & 0.0163 & 0.050 & 99.89 & 61.4k & 96.4k \\
SDBF        & 49.36 & $\leq 0.0022$ & 0.0000 & 100.00 & $\geq 450$k & $\geq 708$k \\
SSF         & 12.99 & 0.0104 & 0.029 & 99.78 & 96.4k & 151.4k \\
PublicCheck & 29.35 & 0.3844 & 0.551 & 98.11 & 2{,}602 & 4{,}082 \\
MiSentry    & 20.80 & 0.6059 & 0.817 & 96.07 & 1{,}651 & 2{,}589 \\
Mismatched (worst case)    & 29.35 & 22.79 & 22.04 & 24.92 & 44 & 66 \\
\midrule
Genuine & 7.095 & 7.011 & 7.010 & 1.20 & 143 & 221 \\
\bottomrule
\end{tabular}
}
\end{table}

Table~\ref{tab:survivors} reports the measured post-filtering flip probability $p_{\mathrm{aft}}$ alongside the idealized $p_{\mathrm{ideal}}$. Results are pooled over nine configurations. Each configuration uses the 30 tampered models from Section~\ref{sec:theory_bound}, with 500 samples per tampered model. Filtering reduces the flip probability by three to four orders of magnitude for IBSF, SDBF, and SSF and by roughly two for PublicCheck and MiSentry. For every framework-specific IVF, the measured $p_{\mathrm{aft}}$ falls below $p_{\mathrm{ideal}}$, meaning the detector therefore removes sensitive fingerprints more often than chance would predict, so the fingerprints that survive are less useful for revealing tampering than the independence assumption predicts. The columns $n_{\mathbb{E}}$ and $n_{95}$ translate these probabilities into cost, giving the number of queries required for ten flips in expectation and with 95\% probability, respectively. Reaching ten flips with 95\% probability requires between $2{,}589$ (MiSentry) and $151{,}400$ (SSF) fingerprint queries. For SDBF, no surviving flip was observed, so we report a lower bound from a one-sided 95\% upper confidence limit on $p_{\mathrm{aft}}$, which yields at least $708$k queries. Genuine inputs, by contrast, are flagged at  $\mathrm{FPR}=1.20\%$ and flip with probability $7.011\%$, reaching the same detection threshold in $221$ queries. After filtering, fingerprint-based tampering detection requires one order of magnitude more queries than using genuine inputs.

Query efficiency is preserved only in the worst case, where a MiSentry-trained detector filters PublicCheck fingerprints. In this setting, observing ten flips requires 44 queries on average and 66 queries at 95\% confidence, approximately 3.35 times fewer than genuine inputs require at the same confidence level. However, such cases are ony a small subset of the total transfer experiments in Figure \ref{fig:transfer_5x5}. In addition, more effective detectors might further reduce this advantage through higher transferability and we leave this to future work. Overall,  submitting additional fingerprints to overcome filtering eliminates the query-efficiency gains of prior IVFs over genuine samples except for cross-IVF settings where detection transfers poorly.

\subsection{Reduced Detectability Lowers Sensitivity}
\label{sec:defense2}

\begin{figure}[hbt]
\centering
\includegraphics[width=0.72\columnwidth]{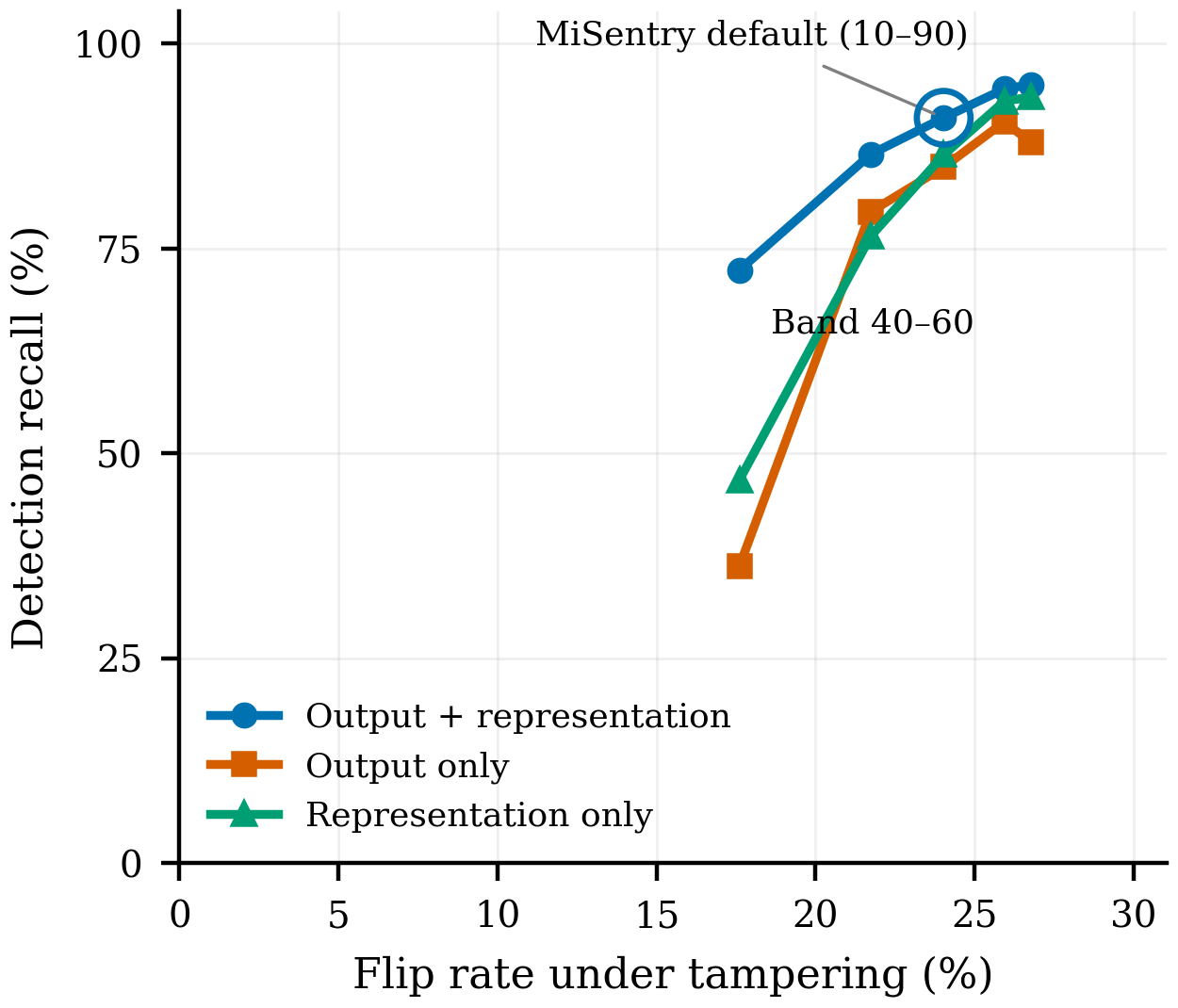}
\caption{Sensitivity--detectability trade-off under MiSentry's own covertness constraint. Each point is one setting of the output-entropy band $[\alpha,\beta]$, swept from the $0$--$100$ to the $40$--$60$ percentile range of normal-sample entropy. Three detectors using only output, only representation, and both features}
\label{fig:misentry_band}
\vspace{-0.4cm}
\end{figure}

\paragraph{Extending MiSentry.}
To recover the query-efficiency advantage lost under framework-specific filtering, we seek to reduce fingerprint detectability while preserving tampering sensitivity. We begin with MiSentry, whose output-entropy interval $[\alpha,\beta]$ provides explicit control over fingerprint concealment. We vary this interval on ResNet-50-ImageNet-1K and measure fingerprint flip rates under $30$ tampered models. We evaluate detectability using three detectors using output features, representation features, or both. As the interval narrows from the $0$--$100$ to the $40$--$60$ percentile range, combined-detector recall decreases from $95.0\%$ to $72.4\%$, but the fingerprint flip rate also decreases from $26.8\%$ to $17.6\%$. The default $10$--$90$ setting provides an intermediate point, with $91.0\%$ recall and a $24.0\%$ flip rate. Thus, tightening the entropy constraint reduces detectability but does not preserve sensitivity. Moreover, detectable signatures remain in both feature groups: at the $40$--$60$ setting, output-only and representation-only detectors achieve $36.2\%$ and $46.7\%$ recall, respectively (Figure~\ref{fig:misentry_band}). The combined detector’s higher recall indicates that representation features carry information complementary to the output features. This motivates concealment constraints that cover both outputs and representations, which we examine next.

\vspace{-0.4cm}
\paragraph{New framework design.} 
Since representation features may still provide strong signals for fingerprint detection besides output statistics, we introduce MAZE (\textbf{M}anifold-\textbf{A}nchored, \textbf{Z}oo-sensitive fingerprints). MAZE constrains fingerprint representations to remain close to the feature distribution of genuine inputs, so they do not look anomalous in representation space (Manifold Constraint). MAZE starts from a pool of genuine sample seeds and tries to make fingerprints sensitive to model modifications by maximizing disagreement between the original model and a zoo of modified models, similar to MiSentry (Zoo sensitivity). The detectors we test against MAZE reuse the corresponding features in each adversarial knowledge setting from Section~\ref{sec:evaluation}. 

\begin{table}[htp]
\centering
\scriptsize
\setlength{\tabcolsep}{2pt}
\caption{MAZE detectability and tampering sensitivity across nine configurations. Rec.\ includes all samples; M.Rec.\ excludes 26.2\% unmodified fingerprints. All-6 includes MAZE in training; LOFO excludes it. $S_F$ measures flips over successfully modified fingerprints and 30 tampered models.}
\label{tab:maze_detection}
\begin{tabular*}{\columnwidth}{@{\extracolsep{\fill}}l|cccc|ccc|ccc|c@{}}
\toprule
& \multicolumn{4}{c|}{(a) Framework-specific} & \multicolumn{3}{c|}{(b) All-6} & \multicolumn{3}{c|}{(c) LOFO} & \\
Dataset & Rec. & M.Rec. & FPR & AUC & Rec. & FPR & AUC & Rec. & FPR & AUC & $S_F$ \\
\midrule
\multicolumn{12}{l}{\textbf{ResNet-50}} \\
CIFAR-100 & 94.0 & 94.4 & 5.7 & .986 & 83.3 & 2.4 & .978 & 58.0 & 2.1 & .909 &  6.98 \\
GTSRB     & 95.6 & 96.4 & 5.3 & .991 & 96.0 & 3.5 & .992 & 60.0 & 3.6 & .962 &  3.06 \\
ImageNet-1K  & 68.4 & 68.4 & 4.5 & .959 & 31.8 & 4.3 & .878 &  2.8 & 3.6 & .601 &  6.87 \\
\midrule
\multicolumn{12}{l}{\textbf{MobileNetV3}} \\
CIFAR-100 & 99.2 & 99.6 & 5.5 & .998 & 97.8 & 3.4 & .995 & 74.8 & 3.3 & .963 & 10.25 \\
GTSRB     & 44.8 & 69.5 & 4.1 & .808 & 42.0 & 4.0 & .704 & 11.0 & 3.4 & .607 &  4.32 \\
ImageNet-1K  & 48.2 & 62.6 & 5.3 & .839 & 11.3 & 3.2 & .732 &  2.5 & 3.1 & .622 & 12.96 \\
\midrule
\multicolumn{12}{l}{\textbf{ViT-B/16}} \\
CIFAR-100 & 23.8 & 56.6 & 5.7 & .659 & 12.0 & 3.7 & .554 &  6.8 & 4.1 & .522 &  7.82 \\
GTSRB     & 29.2 & 66.7 & 5.7 & .709 & 21.0 & 4.1 & .600 &  6.5 & 4.1 & .526 &  4.01 \\
ImageNet-1K  & 41.0 & 73.7 & 4.9 & .770 & 23.5 & 4.4 & .677 &  4.0 & 3.8 & .589 &  8.42 \\
\midrule
\textbf{Average} & \textbf{60.5} & \textbf{76.4} & \textbf{5.2} & \textbf{.858} & \textbf{46.5} & \textbf{3.7} & \textbf{.790} & \textbf{25.2} & \textbf{3.5} & \textbf{.700} & \textbf{7.19} \\
\bottomrule
\end{tabular*}
\end{table}

Table~\ref{tab:maze_detection} reports MAZE fingerprint detectability. The framework-specific detector achieves 60.47\% average recall at 5.2\% FPR (AUC 0.858), substantially below the 96.1--100\% recall observed for existing IVFs. All-6 ensemble trains a detector using samples from MAZE and the other 5 evaluated IVFs. LOFO detector removes the MAZE samples from training. Recall drops to 46.5\% under All-6 ensemble detection and 25.2\% under MAZE excluded LOFO.

However, these results include fingerprint samples MAZE failed to modify under its constraints. Such fallbacks account for 26.2\% (1{,}178/4{,}500) of the evaluated queries. Excluding them raises framework-specific recall to 76.4\%, showing that unsuccessful generation explains part of the lower detectability. MiSentry modifies nearly all inputs under the same perturbation budget and zoo construction and leads to higher detection results. This suggests that MAZE's additional constraints make successful fingerprint generation more difficult.

MAZE's improved stealth also significantly reduces its tampering sensitivity. Because of the optimization constraints, MAZE accepts only candidates whose prediction margin exceeds the 10th percentile of genuine inputs, excluding many low-margin samples that are more likely to change labels after model modification. Consistent with this restriction, the fingerprints that MAZE successfully modifies achieve an average flip rate of $S_F = 7.19\%$, only slightly above the 7.095\% of genuine inputs. Observing ten flips with these fingerprints therefore requires $n_E = 140$ samples on average and $n_{95} = 216$ samples at 95\% confidence, only marginally fewer than genuine inputs require. MAZE thus lowers detectability at the cost of generation success and tampering sensitivity, and its sensitivity approaches that of genuine inputs. This trade-off is consistent with Theorem~\ref{thm:bound}: since a larger sensitivity gap raises the lower bound on optimal BA, indistinguishable fingerprints must have limited excess sensitivity.


\vspace{-0.4cm}
\paragraph{Scope and discussion.}
Our findings do not imply that all black-box integrity verification methods are insecure. Under framework-specific filtering, all five evaluated IVFs lose their query-efficiency advantage over genuine inputs. However, this outcome depends on detector effectiveness: under the least effective cross-IVF filtering scenario, PublicCheck retains its advantage, requiring only 66 queries compared with 221 genuine queries at 95\% probability. Detectability also extends beyond what sensitivity alone explains: SSF has the smallest estimated sensitivity gap yet remains among the most detectable IVFs, suggesting that much of the signal our detectors exploit stems from generation artifacts rather than sensitivity itself. Theorem~1 thus guarantees a floor on distinguishability, while generation artifacts account for much of the practical detectability. Conversely, MAZE demonstrates that reducing detectability does not necessarily improve verification effectiveness, as its lower detection rates come at the cost of generation success and tampering sensitivity. Our conclusions are limited to the evaluated IVFs, model architectures, datasets, tampering scenarios, and their close variants. Moreover, our detectors ignore metadata such as query timing or account history, which real providers could also exploit. These findings highlight the need to evaluate future IVFs not only by their sensitivity but also by their detectability, generation reliability, and effectiveness against adaptive providers.

\vspace{-0.2cm}
\section{Related Work}
\label{sec:related_work}

Sensitive sample-based IVFs detect model tampering through prediction changes on carefully constructed queries. SSF~\cite{he2019sensitive} optimizes parameter change sensitivity, while AID~\cite{aramoon2021aid} and BCD~\cite{kuttichira2022verification} identify sensitive samples using decision-boundary edge-points and Bayesian optimization, respectively. IBSF~\cite{xiaofan2024intersecting} and SDBF~\cite{bai2025sdbf} exploit multiclass decision boundaries, PublicCheck~\cite{wang2023publiccheck} generates boundary-sensitive inputs using generative models, and MiSentry~\cite{he2024towards} maximizes prediction disagreement with modified model variants. ESF~\cite{bai2025esf} and RESF~\cite{hu2025resf} extend fingerprinting to LLM integrity verification. While these methods emphasize sensitivity, verification efficiency, and aspects of covertness, we investigate whether dishonest providers can distinguish verification queries from genuine inputs.

Prior work on model watermarking examines adversaries that evade ownership verification by recognizing or modifying verification queries~\cite{hitaj2018have,namba2019robust}. Unlike watermarking, which primarily seeks to preserve ownership signals against model modifications and removal attacks, sensitive fingerprint-based integrity verification deliberately constructs inputs whose predictions change under tampering. Existing IVFs also consider covertness, including output-entropy constraints in MiSentry and output naturalness and fingerprint leakage in SDBF. In contrast, we systematically investigate whether the sensitivity that enables fingerprints to detect model tampering also makes them distinguishable from genuine inputs, allowing dishonest providers to identify and filter verification queries.

Prior anomaly-detection research identifies softmax confidence~\cite{hendrycks2016baseline} and class-conditional Mahalanobis distances in feature space~\cite{lee2018simple} as useful detection signals. We investigate whether fingerprint generation introduces similar statistical signatures as these works that dishonest providers can exploit to selectively conceal model tampering.


\section{Conclusion}
We show that the sensitivity advantage of verification fingerprints can also help dishonest providers recognize verification queries and conceal model modifications. Under framework-specific filtering, all five evaluated IVFs lose their query-efficiency advantage over genuine inputs. Attempts to improve fingerprint concealment reduce detectability but incur losses in tampering sensitivity or generation success. These findings expose a sensitivity--detectability tension and show that sensitivity alone does not establish effective verification against adaptive providers.

{
    \small
    \bibliographystyle{ieeenat_fullname}
    \bibliography{reference}
}


\appendix
\setcounter{page}{1}
\setcounter{section}{0}
\renewcommand{\thesection}{\Alph{section}}
\setcounter{table}{0}
\renewcommand{\thetable}{S\arabic{table}}
\setcounter{figure}{0}
\renewcommand{\thefigure}{S\arabic{figure}}


\section{Sensitivity--Detectability Analysis: Proof and Details}
\label{app:proof}
\subsection{Proof of Theorem~\ref{thm:bound}}

We analyze the theoretical relationship between fingerprint sensitivity to model modifications and fingerprint detectability. IVFs rely on fingerprints whose predictions change under model tampering. However, not every modification causes a fingerprint's prediction to change. We therefore define sensitivity as the probability of a prediction change under a distribution of model modifications.

Our analysis builds on two observations. First, genuine inputs may also exhibit prediction changes under tampering, but these changes are generally insufficient for reliable verification with few queries. Second, IVFs improve verification efficiency by generating fingerprints that are more sensitive to model modifications than genuine inputs. We show that this excess sensitivity also creates a distinguishing signal that a dishonest provider can exploit to identify verification queries.

In practice, verification queries are expected to constitute a small fraction of the overall query traffic. We therefore use balanced accuracy to evaluate distinguishability, as it assigns equal weight to fingerprint and genuine-input detection, regardless of their relative prevalence.

Let $P_F$ and $P_G$ denote the fingerprint and genuine-input distributions, respectively, and let $T$ be a distribution over modified model parameters $\theta'$. Given the original model's top-1 prediction $\hat{y}_\theta(x)$, we define the tampering sensitivity of an input $s_T(x)$ and the excess sensitivity of fingerprints $\Delta_T$ as

\begin{equation}
\begin{aligned}
s_T(x) &= \Pr_{\theta'\sim T} [\hat y_{\theta'}(x)\neq\hat y_\theta(x)], \\
\Delta_T &= \mathbb{E}_{P_F}[s_T(x)] - \mathbb{E}_{P_G}[s_T(x)].
\end{aligned}
\label{eq:main_sensitivity}
\end{equation}

\vspace{5pt}

A positive $\Delta_T$ indicates that fingerprints are more sensitive to tampering than genuine inputs. This difference also implies statistical distinguishability: if the two input distributions were identical, their expected sensitivities would be equal. We formalize this relationship in the following theorem.

\begin{theorem}[Sensitivity--Detectability Lower Bound]
\label{thm:app_bound}
For any tamper distribution $T$,
\begin{equation}
\begin{aligned}
\mathrm{TV}(P_F,P_G) &\geq |\Delta_T|, \quad
\mathrm{BA}_{\mathrm{opt}} &\geq
\frac{1+|\Delta_T|}{2},
\end{aligned}
\label{eq:main_bound}
\end{equation}
where $\mathrm{TV}$ denotes total variation distance and $\mathrm{BA}_{\mathrm{opt}}$ is the optimal balanced accuracy for distinguishing fingerprints from genuine inputs. 
\end{theorem}

\begin{proof}
By the variational characterization of total variation~\cite{tsybakov2009introduction}, any measurable function $h:\mathcal{X}\to[0,1]$ satisfies
\[
\left|\mathbb{E}_{P_F}[h(x)]-\mathbb{E}_{P_G}[h(x)]\right|
\leq \mathrm{TV}(P_F,P_G).
\]
Since $s_T(x)\in[0,1]$, choosing $h(x)=s_T(x)$ gives $\mathrm{TV}(P_F,P_G)\geq|\Delta_T|$.

\vspace{5pt}

By the Neyman--Pearson lemma~\cite{neyman1933ix}, the likelihood-ratio test is optimal for distinguishing two distributions. Under equal class priors, its balanced accuracy is $\mathrm{BA}_{\mathrm{opt}}=(1+\mathrm{TV}(P_F,P_G))/2$. Combining this with the first inequality gives us $\mathrm{BA}_{\mathrm{opt}}\geq(1+|\Delta_T|)/2$.
\end{proof}

\subsection{Constructing a Detector from Sensitivity}
\label{app:sensitivity_detector}

The sensitivity gap in Theorem~\ref{thm:bound} provides a signal that a dishonest provider can exploit to distinguish fingerprints from genuine inputs. The central idea is to estimate how frequently an input's predicted label changes under model modifications relative to the original model. When fingerprints exhibit greater average sensitivity than genuine inputs under a given modification distribution, this difference can be exploited to identify verification queries. In particular, inputs exhibiting higher label-flip probabilities can be classified as fingerprints using a suitable detection threshold.

To construct such a detector, consider a collection of $M$ modified models $\{\theta'_1,\ldots,\theta'_M\}$ sampled from the modification distribution $\mathcal{T}$. For a query $x$, the provider estimates its sensitivity as

\begin{equation}
\widehat{s}_{\mathcal{T}}(x) = \frac{1}{M} \sum_{j=1}^{M} \mathbb{I} \left[ \hat{y}_{\theta'_j}(x) \neq \hat{y}_{\theta}(x) \right].
\label{eq:empirical_sensitivity}
\end{equation}

Here, $\widehat{s}_{\mathcal{T}}(x)$ is the fraction of modified models whose predictions disagree with the original model. Assuming we have a threshold $\tau$, a sensitivity-based detector can then classify the query using $\tau$:

\begin{equation}
D_{\tau}(x) = \begin{cases} 1, & \widehat{s}_{\mathcal{T}}(x)\geq\tau,\\ 0, & \text{otherwise}, \end{cases} \label{eq:sensitivity_detector}
\end{equation}

where $D_{\tau}(x)=1$ indicates a fingerprint and $D_{\tau}(x)=0$ indicates a genuine input.

The true-positive rate (TPR) and false-positive rate (FPR) of this detector depend on the threshold $\tau$:
\begin{equation}
\begin{aligned}
\mathrm{TPR}(\tau) &= \Pr_{x\sim P_F}[D_{\tau}(x)=1],\\
\mathrm{FPR}(\tau) &= \Pr_{x\sim P_G}[D_{\tau}(x)=1].
\end{aligned}
\end{equation}
The resulting balanced accuracy is
\begin{equation}
\mathrm{BA}(\tau) = \frac{1+\mathrm{TPR}(\tau)-\mathrm{FPR}(\tau)}{2}.
\end{equation}

We now relate the sensitivity gap to the detector's performance. Since sensitivity is estimated using $M$ modified models, it takes values in $\{0,1/M,\ldots,1\}$. Let $k\in\{1,\ldots,M\}$ index the possible nonzero thresholds. The sensitivity score can be expressed as
\begin{equation}
\widehat{s}_{\mathcal{T}}(x) = \frac{1}{M}\sum_{k=1}^{M}\mathbb{I}\left[\widehat{s}_{\mathcal{T}}(x)\geq\frac{k}{M}\right].
\end{equation}

Because each modified model either agrees or disagrees with the original model's prediction, each modified model gives a binary decision. $\widehat{s}_{\mathcal{T}}(x)$ expresses sensitivity as the average of the detector's binary decisions across all thresholds. For example, an input that changes labels under 18 of 30 modified models is classified as a fingerprint at 18 thresholds, giving a sensitivity score of $0.6$.

The sensitivity gap under the empirical modification distribution is therefore
\begin{equation}
\widehat{\Delta}_{\mathcal{T}} = \frac{1}{M}\sum_{k=1}^{M}\left[\mathrm{TPR}\left(\frac{k}{M}\right)-\mathrm{FPR}\left(\frac{k}{M}\right)\right].
\end{equation}
Thus, the sensitivity gap equals the average difference between the detector's true-positive and false-positive rates across thresholds. When $\widehat{\Delta}_{\mathcal{T}}>0$, at least one threshold $\tau^*$ must achieve a difference no smaller than this average. Consequently,
\begin{equation}
\mathrm{BA}(\tau^*) \geq \frac{1+\widehat{\Delta}_{\mathcal{T}}}{2}.
\label{eq:sensitivity_detector_ba}
\end{equation}
For a negative gap, reversing the detector's decisions gives the corresponding bound using $|\widehat{\Delta}_{\mathcal{T}}|$.

This establishes that excess sensitivity provides an exploitable detection signal, although the guarantee applies to a suitable threshold rather than every threshold. In our analysis, sensitivity is estimated using 30 modified models from six modification families. This detector design is different than the detectors that we discuss for our evaluation.

\begin{table}[h]
\centering
\scriptsize
\setlength{\tabcolsep}{3pt}
\renewcommand{\arraystretch}{1.1}
\caption{Pre-filtering label-flip rates and sensitivity gaps across six model-modification families, averaged over nine model--dataset configurations. Each family contains five modified models per configuration. Flip rates are reported in percentages, and sensitivity gaps in percentage points. PC: PublicCheck; MS: MiSentry.}
\label{tab:modification_flip_rates}
\resizebox{\columnwidth}{!}{
\begin{tabular}{lrrrrrr}
\toprule
Model modification & IBSF & SDBF & SSF & PC & MS & Genuine \\
\midrule
\multicolumn{7}{l}{\textbf{Label-flip rate (\%)}} \\
Last-layer fine-tuning & 15.47 & 21.36 & 0.01 & 2.46 & 0.50 & 0.24 \\
Targeted attack        & 13.19 & 18.95 & 18.88 & 5.61 & 4.48 & 0.86 \\
BadNets backdoor       & 76.25 & 79.42 & 30.61 & 49.76 & 38.71 & 6.44 \\
Bit-flip attack        & 47.24 & 48.01 & 0.03 & 25.74 & 10.02 & 2.33 \\
Weight quantization    & 61.04 & 66.39 & 18.76 & 51.72 & 43.02 & 21.09 \\
Magnitude pruning      & 61.39 & 62.04 & 9.68 & 40.84 & 28.06 & 11.61 \\
\midrule
\textbf{Average} & \textbf{45.76} & \textbf{49.36} & \textbf{12.99} & \textbf{29.35} & \textbf{20.80} & \textbf{7.09} \\
\midrule
\multicolumn{7}{l}{\textbf{Sensitivity gap $\widehat{\Delta}_{\mathcal{T}_j}$ (percentage points)}} \\
Last-layer fine-tuning & 15.23 & 21.12 & -0.23 & 2.22 & 0.26 & -- \\
Targeted attack        & 12.33 & 18.09 & 18.02 & 4.75 & 3.62 & -- \\
BadNets backdoor       & 69.81 & 72.98 & 24.17 & 43.32 & 32.27 & -- \\
Bit-flip attack        & 44.91 & 45.68 & -2.30 & 23.41 & 7.69 & -- \\
Weight quantization    & 39.95 & 45.30 & -2.33 & 30.63 & 21.93 & -- \\
Magnitude pruning      & 49.78 & 50.43 & -1.93 & 29.23 & 16.45 & -- \\
\midrule
\textbf{Average} & \textbf{38.67} & \textbf{42.27} & \textbf{5.90} & \textbf{22.26} & \textbf{13.71} & -- \\
\bottomrule
\end{tabular}
}
\end{table}

\subsection{Model Modification Distribution}
\label{app:model_modifications}

We generate 30 modified models per model--dataset configuration, with five variants from each of six modification families, resulting in 270 models across nine configurations.

\paragraph{Last-layer fine-tuning.} We fine-tune the classification head using AdamW ($10^{-5}$) until a probe prediction changes or 5,000 steps are reached.

\paragraph{Targeted attack.} We select a correctly classified sample and fine-tune the source and target class parameters to redirect its prediction toward the nearest competing class centroid in feature space, while preserving predictions on other samples.

\paragraph{BadNets backdoor.} We poison 10\% of training samples with a $16\times16$ checkerboard trigger and fine-tune the full model using SGD ($10^{-3}$), targeting 90\% attack success with at most a 3-percentage-point clean-accuracy drop.

\paragraph{Bit-flip attack.} We apply gradient-guided bit flips to an INT8-quantized model, prioritizing changes that alter the fewest predictions on a probe batch, with at most five flips.

\paragraph{Weight quantization.} We apply per-channel weight-only quantization at 8, 7, 6, 5, and 4 bits.

\paragraph{Magnitude pruning.} We globally prune the smallest-magnitude weights at rates of 5\%, 10\%, 20\%, 30\%, and 40\%.

For training-based modifications, we use 5,000 training samples and 2,000 for evaluation. Fingerprint source images are excluded from these sets. The empirical modification distribution $\widehat{\mathcal{T}}$ assigns equal weight to all 30 models per configuration.

\subsection{Empirical Estimation of the Sensitivity Gap}

Table~\ref{tab:modification_flip_rates} reports the label-flip probabilities of fingerprints and genuine inputs across six model-modification families. Each value is averaged over nine model--dataset configurations and five modified models per family. Averaging over all six families yields the empirical sensitivity gaps reported in Table~1 of the main paper. Here, we express sensitivity gaps in percentage points to support direct comparison between fingerprint and genuine-input flip rates, whereas the main paper reports these gaps in decimal form to make it easier to explain detectability lower bounds.


\section{Detector Features}
\label{app:features}

Table~\ref{tab:feature_comparison} summarizes the features used by our framework-specific and ensemble detectors. Many of these features have been previously studied for out-of-distribution (OOD) and generated-image detection~\cite{hendrycks2016baseline,hendrycks2019scaling,liu2020energy,lee2018simple,sun2022out,huang2021importance,park2023understanding,frank2020leveraging}. We adapt these features to distinguish fingerprints from genuine inputs and organize them into four categories.

\paragraph{Output statistics.} Output statistics are useful to characterize the model's prediction confidence, probability distribution, and logit patterns to identify differences between fingerprints and genuine inputs. These features help to identify the model inference-based distinctions between fingerprints and genuine samples.

\paragraph{Representation statistics.} Representation statistics measure how a sample’s intermediate model features differ from those of genuine inputs using distance-based and reconstruction-based measures. These features help us to measure the differences in intermediate representations, which might not always be present in the output statistics.

\paragraph{Closed-form sensitivity measures.} Closed-form sensitivity measures estimate how sensitive model predictions are to small parameter changes using analytical calculations without additional model evaluations. These features provide analytical measures of local output sensitivity.

\begin{table}[h]
\centering
\scriptsize
\setlength{\tabcolsep}{2pt}
\renewcommand{\arraystretch}{1.05}
\caption{Feature categories used by ensemble and framework-specific detectors. EN: Ensemble; IB: IBSF; SD: SDBF; SS: SSF; PC: PublicCheck; MS: MiSentry; MZ: MAZE.}
\label{tab:feature_comparison}
\resizebox{\columnwidth}{!}{
\begin{tabular}{lccccccc}
\toprule
Feature category & EN & IB & SD & SS & PC & MS & MZ \\
\midrule
\multicolumn{8}{l}{\textbf{Output Statistics}} \\
Entropy                       & \checkmark & \checkmark & \checkmark & \checkmark & \checkmark & \checkmark & \checkmark \\
Effective number of classes   & \checkmark & \checkmark & \checkmark & \checkmark & \checkmark & \checkmark & \checkmark \\
Confidence statistics         & \checkmark & \checkmark & \checkmark & \checkmark & \checkmark & \checkmark & \checkmark \\
Probability margins           & \checkmark & \checkmark & \checkmark & \checkmark & \checkmark & \checkmark & \checkmark \\
Top-$k$ probability stats.    & \checkmark & \checkmark & \checkmark & \checkmark & \checkmark & \checkmark & \checkmark \\
Partial/normalized entropy    & \checkmark & \checkmark & \checkmark & \checkmark & -- & -- & \checkmark \\
Top-$k$ uniformity            & \checkmark & \checkmark & \checkmark & \checkmark & -- & -- & \checkmark \\
Logit statistics              & \checkmark & \checkmark & \checkmark & \checkmark & \checkmark & \checkmark & \checkmark \\
Logit boundary/cliff          & \checkmark & \checkmark & \checkmark & \checkmark & -- & -- & \checkmark \\
Top-$K$ logit variance        & -- & -- & \checkmark & -- & -- & -- & \checkmark \\
SSF sensitivity objective     & -- & -- & -- & \checkmark & -- & -- & -- \\
Last-layer gradient stats.    & -- & -- & -- & \checkmark & -- & -- & -- \\
Penultimate activation stats. & -- & -- & -- & \checkmark & -- & -- & -- \\
Entropy-band margin           & -- & -- & -- & -- & -- & \checkmark & \checkmark \\
\midrule
\multicolumn{8}{l}{\textbf{Representation Statistics}} \\
Class-cond. Mahalanobis       & \checkmark & \checkmark & \checkmark & \checkmark & \checkmark & \checkmark & \checkmark \\
Pred.-class Mahalanobis       & \checkmark & \checkmark & \checkmark & \checkmark & \checkmark & \checkmark & \checkmark \\
Mahalanobis class gap         & \checkmark & \checkmark & \checkmark & \checkmark & \checkmark & \checkmark & \checkmark \\
Global Mahalanobis            & \checkmark & \checkmark & \checkmark & \checkmark & \checkmark & \checkmark & \checkmark \\
Mean $k$-NN distance          & \checkmark & \checkmark & \checkmark & \checkmark & \checkmark & \checkmark & \checkmark \\
Nearest-neighbor distance     & \checkmark & \checkmark & \checkmark & \checkmark & \checkmark & \checkmark & \checkmark \\
PCA reconstruction residual   & \checkmark & \checkmark & \checkmark & \checkmark & \checkmark & \checkmark & \checkmark \\
Multi-depth representations   & \checkmark & -- & \checkmark & -- & -- & -- & \checkmark \\
\midrule

\multicolumn{8}{l}{\textbf{Closed-Form Sensitivity}} \\
Softmax Jacobian norm         & \checkmark & -- & \checkmark & -- & -- & -- & \checkmark \\
Log-Jacobian norm             & \checkmark & -- & -- & -- & -- & -- & -- \\
Penultimate activation norm   & \checkmark & -- & -- & \checkmark & -- & -- & -- \\
\midrule
\multicolumn{8}{l}{\textbf{Image Statistics}} \\
Total variation               & \checkmark & -- & -- & -- & \checkmark & -- & -- \\
Laplacian energy              & \checkmark & -- & -- & -- & \checkmark & -- & -- \\
Block-boundary contrast       & \checkmark & -- & -- & -- & \checkmark & -- & -- \\
Spectral energy               & \checkmark & -- & -- & -- & \checkmark & -- & -- \\
Spectral periodicity          & \checkmark & -- & -- & -- & \checkmark & -- & -- \\
Pixel intensity statistics    & \checkmark & -- & -- & -- & \checkmark & -- & -- \\
Saturation statistics         & \checkmark & -- & -- & -- & \checkmark & -- & -- \\
Colorfulness                  & \checkmark & -- & -- & -- & \checkmark & -- & -- \\
Unique intensity levels       & \checkmark & -- & -- & -- & \checkmark & -- & -- \\
\bottomrule
\end{tabular}
}
\end{table}

\begin{table*}[!t]
\centering
\footnotesize
\setlength{\tabcolsep}{3pt}
\renewcommand{\arraystretch}{1.05}
\caption{Framework-specific fingerprint detection across nine model--dataset configurations.}
\label{tab:framework_specific}
\begin{tabular*}{\textwidth}{@{\extracolsep{\fill}}ll cc cc cc cc cc}
\toprule
\multirow{2}{*}{Model} & \multirow{2}{*}{Dataset}
 & \multicolumn{2}{c}{IBSF} & \multicolumn{2}{c}{SDBF} & \multicolumn{2}{c}{SSF}
 & \multicolumn{2}{c}{PublicCheck} & \multicolumn{2}{c}{MiSentry} \\
\cmidrule(lr){3-4}\cmidrule(lr){5-6}\cmidrule(lr){7-8}\cmidrule(lr){9-10}\cmidrule(lr){11-12}
 & & Rec./FPR & BA/AUC & Rec./FPR & BA/AUC & Rec./FPR & BA/AUC & Rec./FPR & BA/AUC & Rec./FPR & BA/AUC \\
\midrule
\multirow{3}{*}{ResNet-50}
 & CIFAR-100   & 99.8/0.1  & 99.9/1.000  & 100.0/0.0 & 100.0/1.000 & 100.0/0.1 & 100.0/1.000 & 98.6/1.0 & 98.8/0.999 & 91.6/3.0 & 94.3/0.991 \\
 & GTSRB       & 100.0/0.0 & 100.0/1.000 & 100.0/0.0 & 100.0/1.000 & 99.8/0.1  & 99.9/1.000  & 96.4/0.1 & 98.2/1.000 & 93.0/8.3 & 92.4/0.978 \\
 & ImageNet-1K & 99.8/0.0  & 99.9/1.000  & 100.0/0.0 & 100.0/1.000 & 99.4/0.0  & 99.7/1.000  & 96.2/0.8 & 97.7/0.999 & 98.2/3.4 & 97.4/0.996 \\
\addlinespace[2pt]
\multirow{3}{*}{MobileNetV3}
 & CIFAR-100   & 99.8/0.0  & 99.9/1.000  & 100.0/0.0 & 100.0/1.000 & 99.6/0.1  & 99.8/1.000  & 98.1/1.5 & 98.3/0.999 & 98.2/6.2 & 96.0/0.993 \\
 & GTSRB       & 100.0/0.0 & 100.0/1.000 & 100.0/0.0 & 100.0/1.000 & 100.0/0.0 & 100.0/1.000 & 98.8/1.2 & 98.8/0.999 & 98.4/3.4 & 97.5/0.996 \\
 & ImageNet-1K & 100.0/0.0 & 100.0/1.000 & 100.0/0.0 & 100.0/1.000 & 100.0/0.0 & 100.0/1.000 & 99.6/1.8 & 98.9/0.999 & 98.8/0.3 & 99.3/1.000 \\
\addlinespace[2pt]
\multirow{3}{*}{ViT-B/16}
 & CIFAR-100   & 99.8/0.0  & 99.9/1.000  & 100.0/0.0 & 100.0/1.000 & 99.6/0.4  & 99.6/1.000  & 98.9/1.4 & 98.8/0.999 & 93.8/9.4 & 92.2/0.969 \\
 & GTSRB       & 100.0/0.0 & 100.0/1.000 & 100.0/0.0 & 100.0/1.000 & 100.0/0.2 & 99.9/1.000  & 98.4/0.7 & 98.9/1.000 & 98.4/2.8 & 97.8/0.997 \\
 & ImageNet-1K & 99.8/0.0  & 99.9/1.000  & 100.0/0.0 & 100.0/1.000 & 99.6/0.0  & 99.8/1.000  & 97.8/0.6 & 98.6/1.000 & 94.2/6.7 & 93.8/0.982 \\
\midrule
\multicolumn{2}{l}{\textbf{Average}}
 & \textbf{99.9/0.0} & \textbf{99.9/1.000} & \textbf{100.0/0.0} & \textbf{100.0/1.000} & \textbf{99.8/0.1} & \textbf{99.8/1.000} & \textbf{98.1/1.0} & \textbf{98.6/0.999} & \textbf{96.1/4.8} & \textbf{95.6/0.989} \\
\bottomrule
\end{tabular*}
\end{table*}

\begin{table*}[!t]
\centering
\footnotesize
\setlength{\tabcolsep}{3pt}
\renewcommand{\arraystretch}{1.05}
\caption{Ensemble fingerprint detection across nine model--dataset configurations trained on all five IVFs.}
\label{tab:ensemble}
\begin{tabular*}{\textwidth}{@{\extracolsep{\fill}}ll cc cc cc cc cc}
\toprule
\multirow{2}{*}{Model} & \multirow{2}{*}{Dataset}
 & \multicolumn{2}{c}{IBSF} & \multicolumn{2}{c}{SDBF} & \multicolumn{2}{c}{SSF}
 & \multicolumn{2}{c}{PublicCheck} & \multicolumn{2}{c}{MiSentry} \\
\cmidrule(lr){3-4}\cmidrule(lr){5-6}\cmidrule(lr){7-8}\cmidrule(lr){9-10}\cmidrule(lr){11-12}
 & & Rec./FPR & BA/AUC & Rec./FPR & BA/AUC & Rec./FPR & BA/AUC & Rec./FPR & BA/AUC & Rec./FPR & BA/AUC \\
\midrule
\multirow{3}{*}{ResNet-50}
 & CIFAR-100 & 100.0/3.0 & 98.5/1.000 & 100.0/3.0 & 98.5/1.000 & 100.0/3.0 & 98.5/1.000 & 100.0/3.0 & 98.5/1.000 & 100.0/3.0 & 98.5/1.000 \\
 & GTSRB & 100.0/2.9 & 98.6/0.999 & 100.0/2.9 & 98.6/0.999 & 100.0/2.9 & 98.6/0.999 & 100.0/2.9 & 98.6/0.999 & 80.8/2.9 & 89.0/0.976 \\
 & ImageNet-1K & 100.0/3.6 & 98.2/1.000 & 100.0/3.6 & 98.2/1.000 & 100.0/3.6 & 98.2/1.000 & 96.7/3.6 & 96.6/0.994 & 97.8/3.6 & 97.1/0.996 \\
\addlinespace[2pt]
\multirow{3}{*}{MobileNetV3}
 & CIFAR-100 & 100.0/3.1 & 98.5/0.999 & 100.0/3.1 & 98.5/0.999 & 100.0/3.1 & 98.5/0.999 & 98.0/3.1 & 97.5/0.996 & 99.3/3.1 & 98.1/0.997 \\
 & GTSRB & 100.0/2.8 & 98.6/1.000 & 100.0/2.8 & 98.6/1.000 & 100.0/2.8 & 98.6/1.000 & 99.9/2.8 & 98.6/0.999 & 97.8/2.8 & 97.5/0.995 \\
 & ImageNet-1K & 100.0/3.3 & 98.4/1.000 & 100.0/3.3 & 98.4/1.000 & 100.0/3.3 & 98.4/1.000 & 96.7/3.3 & 96.7/0.995 & 100.0/3.3 & 98.4/1.000 \\
\addlinespace[2pt]
\multirow{3}{*}{ViT-B/16}
 & CIFAR-100 & 100.0/3.0 & 98.5/0.999 & 100.0/3.0 & 98.5/0.999 & 100.0/3.0 & 98.5/0.999 & 98.5/3.0 & 97.7/0.996 & 75.5/3.0 & 86.3/0.966 \\
 & GTSRB & 100.0/3.0 & 98.5/1.000 & 100.0/3.0 & 98.5/1.000 & 100.0/3.0 & 98.5/1.000 & 100.0/3.0 & 98.5/0.999 & 98.0/3.0 & 97.5/0.997 \\
 & ImageNet-1K & 100.0/5.7 & 97.2/0.998 & 100.0/5.7 & 97.2/0.998 & 100.0/5.7 & 97.2/0.998 & 99.0/5.7 & 96.7/0.993 & 67.8/5.7 & 81.1/0.947 \\
\midrule
\multicolumn{2}{l}{\textbf{Average}}
 & \textbf{100.0/3.3} & \textbf{98.3/0.999}
 & \textbf{100.0/3.3} & \textbf{98.3/0.999}
 & \textbf{100.0/3.3} & \textbf{98.3/0.999}
 & \textbf{98.7/3.3} & \textbf{97.7/0.997}
 & \textbf{90.8/3.3} & \textbf{93.7/0.986} \\
\bottomrule
\end{tabular*}
\end{table*}

\begin{table*}[!t]
\centering
\footnotesize
\setlength{\tabcolsep}{3pt}
\renewcommand{\arraystretch}{1.05}
\caption{Leave-one-framework-out (LOFO) fingerprint detection across nine model--dataset configurations. Each ensemble is trained on the other four IVFs and evaluated on the held-out IVF.}
\label{tab:ensemble_lofo}
\begin{tabular*}{\textwidth}{@{\extracolsep{\fill}}ll cc cc cc cc cc}
\toprule
\multirow{2}{*}{Model} & \multirow{2}{*}{Dataset}
 & \multicolumn{2}{c}{IBSF} & \multicolumn{2}{c}{SDBF} & \multicolumn{2}{c}{SSF}
 & \multicolumn{2}{c}{PublicCheck} & \multicolumn{2}{c}{MiSentry} \\
\cmidrule(lr){3-4}\cmidrule(lr){5-6}\cmidrule(lr){7-8}\cmidrule(lr){9-10}\cmidrule(lr){11-12}
 & & Rec./FPR & BA/AUC & Rec./FPR & BA/AUC & Rec./FPR & BA/AUC & Rec./FPR & BA/AUC & Rec./FPR & BA/AUC \\
\midrule
\multirow{3}{*}{ResNet-50}
 & CIFAR-100 & 100.0/4.1 & 98.0/1.000 & 100.0/3.4 & 98.3/1.000 & 100.0/4.0 & 98.0/1.000 & 82.5/3.1 & 89.7/0.987 & 100.0/3.3 & 98.4/0.998 \\
 & GTSRB & 100.0/3.3 & 98.4/1.000 & 100.0/3.5 & 98.3/0.999 & 100.0/3.4 & 98.3/0.999 & 94.0/2.9 & 95.6/0.987 & 46.0/2.4 & 71.8/0.873 \\
 & ImageNet-1K & 100.0/4.0 & 98.0/1.000 & 100.0/4.3 & 97.9/1.000 & 100.0/4.0 & 98.0/1.000 & 11.7/3.2 & 54.2/0.740 & 89.3/3.7 & 92.8/0.969 \\
\addlinespace[2pt]
\multirow{3}{*}{MobileNetV3}
 & CIFAR-100 & 100.0/3.5 & 98.3/0.999 & 100.0/3.3 & 98.4/0.999 & 100.0/3.1 & 98.5/0.999 & 65.6/3.0 & 81.3/0.960 & 92.3/2.9 & 94.7/0.989 \\
 & GTSRB & 100.0/2.6 & 98.7/1.000 & 100.0/2.9 & 98.6/1.000 & 100.0/3.1 & 98.5/1.000 & 84.4/2.7 & 90.9/0.985 & 73.0/2.3 & 85.4/0.967 \\
 & ImageNet-1K & 100.0/3.3 & 98.4/1.000 & 100.0/3.6 & 98.2/1.000 & 100.0/3.2 & 98.4/1.000 & 2.1/2.5 & 49.8/0.605 & 99.8/3.1 & 98.4/0.999 \\
\addlinespace[2pt]
\multirow{3}{*}{ViT-B/16}
 & CIFAR-100 & 100.0/3.6 & 98.2/0.999 & 100.0/3.7 & 98.2/0.999 & 100.0/3.1 & 98.5/0.994 & 41.0/3.1 & 69.0/0.879 & 1.8/2.7 & 49.5/0.802 \\
 & GTSRB & 100.0/3.6 & 98.2/0.999 & 100.0/3.4 & 98.3/1.000 & 99.3/2.9 & 98.2/0.998 & 92.2/3.1 & 94.5/0.990 & 71.0/2.8 & 84.1/0.965 \\
 & ImageNet-1K & 100.0/6.0 & 97.0/0.998 & 100.0/6.2 & 96.9/0.998 & 83.0/5.6 & 88.7/0.935 & 16.1/4.7 & 55.7/0.870 & 18.5/5.3 & 56.6/0.856 \\
\midrule
\multicolumn{2}{l}{\textbf{Average}}
 & \textbf{100.0/3.8} & \textbf{98.1/0.999}
 & \textbf{100.0/3.8} & \textbf{98.1/0.999}
 & \textbf{98.0/3.6} & \textbf{97.2/0.992}
 & \textbf{54.4/3.1} & \textbf{75.6/0.889}
 & \textbf{65.7/3.1} & \textbf{81.3/0.935} \\
\bottomrule
\end{tabular*}
\end{table*}

\paragraph{Image statistics.} Image statistics characterize spatial, spectral, and pixel-level properties of inputs to identify artifacts introduced during fingerprint generation. These features are particularly relevant to PublicCheck, whose latent-space generation process may introduce image artifacts that distinguish fingerprints from genuine samples.

Framework-specific detectors combine general features with those motivated by their fingerprint-generation objectives. For example, SDBF uses top-$K$ logit variance, SSF incorporates last-layer sensitivity features, and MiSentry uses entropy-band margins. The ensemble excludes IVF-specific features while retaining general features across all four categories, allowing detection without prior knowledge of the deployed IVF. MAZE's detector additionally measures conformity to MAZE's own logit-margin and Mahalanobis-radius constraints. All features are extracted from a single model forward pass, using precomputed reference statistics generated from the training phase.


\section{Evaluation}
\label{app:evaluation}

This section provides additional details on the experimental setup, hyperparameters, and fingerprint generation procedures.

\subsection{Extended Experimental Setup}
\label{app:setup}

We evaluate five IVFs (IBSF, SDBF, SSF, PublicCheck, and MiSentry) under three adversarial knowledge settings.

\paragraph{Framework-specific detection.} The adversary knows the deployed fingerprint-generation algorithm and can independently generate fingerprints for detector training. This assumption is consistent with the threat models of existing IVFs.

\paragraph{Ensemble detection.} The adversary does not know which IVF is deployed, but fingerprints from the target IVF are included in the detector's training data. We use independent fingerprint-generation runs for evaluation to ensure that training and test samples are disjoint.

\paragraph{Cross-framework generalization.} The adversary does not know the deployed IVF, and its fingerprints are excluded from detector training. We evaluate whether an ensemble trained on the remaining IVFs can identify fingerprints from the unseen framework.

We evaluate nine model--dataset configurations, combining ResNet-50~\cite{resnet50}, MobileNetV3~\cite{mobilenetv3}, and ViT-B/16~\cite{vit_b16} with CIFAR-100~\cite{cifar100}, GTSRB~\cite{gtsrb}, and ImageNet-1K~\cite{imagenet} under a standardized evaluation pipeline.

\paragraph{Fingerprint Generation.} We use the official IBSF implementation, adapt the released SSF and PublicCheck code~\cite{ssf_code,publiccheck_code}, and reimplement SDBF and MiSentry. All generators use dataset-specific normalization and 8-bit image quantization. The four pixel-space IVFs enforce $\|\delta\|_2\leq5$, measured over the full image in $[0,1]$ pixel space before channel normalization. PublicCheck operates in latent space without an enforced pixel-space perturbation budget. SDBF uses $K=10$ and 5{,}000 optimization steps, similar to IBSF, while MiSentry uses 12 modified models per configuration.

We report recall, false-positive rate (FPR), balanced accuracy (BA), and AUC. We emphasize recall because each correctly identified fingerprint allows the adversary to return the original model's prediction, potentially concealing tampering.

\subsection{Framework-Specific Detectors}
\label{app:eval_specific}

Table~\ref{tab:framework_specific} presents the complete framework-specific detection results across nine model--dataset configurations. Each reported value is averaged over five independent runs.


\subsection{Ensemble Detectors}
\label{app:eval_ensemble}

Table~\ref{tab:ensemble} presents the complete ensemble detection results across nine model--dataset configurations. For each configuration, we train the ensemble detector using fingerprints from all five IVFs and evaluate it on independently generated fingerprints from each target IVF. Each reported value is averaged over five independent runs of this evaluation pipeline.


\subsection{Detector Generalization}
\label{app:eval_generalization}

Table~\ref{tab:ensemble_lofo} presents the complete leave-one-framework-out (LOFO) detection results across nine model--dataset configurations. For each configuration, we train the ensemble detector using fingerprints from four IVFs and evaluate it on independently generated fingerprints from the held-out IVF, which is excluded from training. Each reported value is averaged over five independent runs of this evaluation pipeline.

\end{document}